\documentclass[aps,pre,floats,nofootinbib,twocolumn,showpacs,superscriptaddress,reprint]{revtex4-1}
\usepackage{graphicx,epsfig}
\usepackage{graphics,dcolumn,bm,epic, eepic,fleqn,float}
\usepackage{amssymb,amsmath,amsfonts,multirow,rotate,color,xcolor}
\usepackage{soul}
\usepackage{url}

\usepackage{subfigure}

\begin{document}
\author{Chlo\"e Brown}
\affiliation{Computer Laboratory, University of Cambridge, Cambridge (UK)}

\author{Christos Efstratiou}
\affiliation{Computer Laboratory, University of Cambridge, Cambridge (UK)}
\affiliation{School of Engineering \& Digital Arts, University of Kent (UK)}

\author{Ilias Leontiadis}
\affiliation{Telefonica Research, Barcelona (Spain)}
\altaffiliation{Data collection was undertaken while the authors were at the Computer Laboratory of the University of Cambridge.}

\author{Daniele Quercia}
\affiliation{Yahoo Labs, Barcelona (Spain)}
\altaffiliation{Data collection was undertaken while the authors were at the Computer Laboratory of the University of Cambridge.}

\author{Cecilia Mascolo}
\affiliation{Computer Laboratory, University of Cambridge, Cambridge (UK)}
\date{\today}

\begin{abstract}In many work environments, serendipitous interactions between members of different groups may lead to enhanced productivity, collaboration and knowledge dissemination. Two factors that may have an influence on such interactions are cultural differences between individuals in highly multicultural workplaces, and the layout and physical spaces of the workplace itself.
In this work, we investigate how these two factors may facilitate or hinder inter-group interactions in the workplace. We analyze traces collected using wearable electronic badges to capture face-to-face interactions and mobility patterns of employees in a research laboratory in the UK. We observe that those who interact with people of different roles tend to come from collectivist cultures that value relationships and where people tend to be comfortable with social hierarchies, and that some locations in particular are more likely to host serendipitous interactions, knowledge that could be used by organizations to enhance communication and productivity.\end{abstract}
\title{Tracking Serendipitous Interactions: How Individual Cultures Shape the Office}
\maketitle
\section{Introduction}
In many knowledge-based work environments, where creativity and innovation are key, it is intuitive that serendipitous interactions between members of different teams, with complementary expertise or skill sets, can be highly beneficial as a source of fresh perspectives, information, and ideas. Chance conversations between individuals who are not necessarily part of the same group have long been judged to be essential for team coordination, cohesiveness and productivity~\cite{burt2004structural, Isaacs96:Piazza, Jeffrey00:Sharing, Whittaker94:Informal}. This idea was recently further confirmed by Pentland \emph{et al.}, in a study of workplace communication patterns at a Prague bank~\cite{Pentland12:New}. They discovered a key characteristic of successful teams: members periodically interact with others outside of their team, and bring back new information. They dubbed this critical dimension of communication `exploration', the tendency for a team to seek inter-group connections, and found that teams with high exploration tended to be more successful, demonstrating the power of serendipitous interactions in the office.

One factor that could clearly affect the ease of such interactions is the physical spaces of the workplace itself; for example, high-traffic areas such as coffee machines and photocopiers may be particularly likely places for inter-group meetings~\cite{fayard2007photocopiers,Isaacs96:Piazza}. In general, if spaces encourage the mixing and meeting of a diverse range of people, serendipitous meetings between individuals from different teams or social groups will occur more readily, which could be crucial; face-to-face communication has been shown to be more important than electronic means such as email or SMS~\cite{Pentland12:New, Stryker12:Facilitating}. 

Indeed, designing the layout of workspaces in accordance with office social dynamics resulting from the background of employees can help them work more effectively, as was shown by a recent study of offices across 11 different countries~\cite{Steelcase13:Culture}. While this approach takes into account the national culture of the country where an office is situated, it does not consider what happens in very culturally diverse environments that accommodate workers from many different backgrounds. 

The effects of cultural variations between countries on organisations have been studied by Geert Hofstede, by means of administering opinion surveys to IBM employees in over 70 countries. He derived five main factors, or cultural dimensions, that can account for most of the variance he observed~\cite{Hofstede10:Cultures}, and found that workers' cultural backgrounds can shape the way they think, feel, and act. Besides the workplace environment itself, cultural differences between its occupants could affect office social dynamics and, therefore, the propensity of workers to engage in beneficial serendipitous inter-group interactions.

Since interactions between people with different areas of expertise or social circles are so beneficial for the exchange of information and ideas, it is important to understand what facilitates or hinders them. Studies of workplace communication have traditionally used pen-and-paper methods from the social sciences, with data being gathered through direct manual observations and participant surveys. These approaches have various disadvantages and limitations: the presence of an observer may cause people to reflect upon and change their natural behavior~\cite{Whyte43:Street}, and surveys can suffer from participants giving answers they feel are socially desirable, or misremembering~\cite{Bradburn78:Question,Van08:Faking}. Recently, technological methods such as wearable badges and sensors have enabled social interaction patterns to be studied in a less obtrusive way~\cite{Olguin09:Sensible,Wu08:Mining}.  In this work, we make use of this technology to investigate the impact of cultural differences on serendipitous interactions in the workplace in a way that was not available to Hofstede doing his original work in the 1970s. 

In addition, we can make use of localization available from the sensing technology to study the characteristics of different spaces within the workplace, and their potential to host encounters between people from different groups, knowledge that could be used by organizations to stimulate inter-group meetings and perhaps enhance productivity.

We present a study of serendipitous interactions in an office environment, making the following contributions:
\begin{itemize}
	\item We analyze a dataset captured using the SocioPatterns proximity-sensing platform \footnote{\texttt{http://www.sociopatterns.org}}. For a period of 2 weeks, 61 people at a research laboratory in the UK wore lightweight electronic badges that can capture face-to-face interactions (as opposed to Bluetooth devices, which simply capture colocation information)~\cite{Cattuto10:Dynamics}. The use of these badges allows us to study face-to-face interactions and mobility in the office environment in a less obtrusive way than older sensing technologies.
	\item We investigate the relationship between Hofstede's cultural dimensions and detected interactions. We analyze Power Distance (how comfortable people are with hierarchy), Individualism (how individualist or collectivist a culture is), and Masculinity (task-orientation vs.~person-orientation), finding that those who interact with people of different roles tend to come from collectivist and person-oriented cultures comfortable with social hierarchies. This suggests that cultural effects depend not only on the location of the office itself, but on the cultural diversity of the people working there.
	\item We assess the potential for particular rooms to host serendipitous interactions between people from different groups, by counting visitors to the room from groups defined according to a variety of dimensions: research group, role, nationality, age, and gender. We determine the extent to which different kinds of spaces allow for serendipitous interactions between visitors, and find that while social spaces such as kitchens and common rooms score highly for all dimensions, others such as shared offices score more highly in some dimensions than others. This means that different locations may be more important for the promotion of interactions between individuals from different groups in the workplace, depending on the kind of diversity it is wished to achieve.
\end{itemize}

Our work demonstrates the feasibility of detecting serendipitous interactions in the workplace using unobtrusive sensing technology. Given the importance of interactions between people from different teams and groups for productivity, information flow, and idea exchange, being able to sense and quantify such effects could potentially be highly beneficial to organizations.

\section{Related work}
\paragraph{Electronic sensing of office social interactions} 
From as early as 1992, the design of the Active Badge~\cite{Want92:ActiveBadge} system was motivated by the need for location-based services in business environments. Most of these early attempts were focused on the design of context-aware systems, placing less emphasis on the potential of using location technologies to understand and analyze social interactions.

Olgu\'in-Olgu\'in \emph{et al.}~\cite{Olguin09:Sensible} demonstrated the feasibility of using wearable computing devices to measure, among a variety of factors, face-to-face interactions in the workplace. They showed that the data collected by the devices could be used in combination with email communication data to predict employees' perceptions of group interaction. Their work differs from the study we present here in that that the sensing devices used were bigger, and thus potentially more obtrusive, than the tiny badges we use. Moreover, they were not looking explicitly to study encounters between individuals from different cultural backgrounds. 

The same technology has been used in a variety of studies of social interactions in the workplace, for example by Waber et. al.~\cite{Waber10:Productivity}, who studied the effect of social group strength on productivity, and by Wu \emph{et al.}~\cite{Wu08:Mining}, who compared face-to-face network structure with the network formed by electronic communication.

Recently, research projects have shown that social interactions in the workplace can be detected using mobile phones~\cite{efs12a}. Although mobile phones are demonstrably less accurate in detecting face-to-face interactions than specialised wearable devices, these attempts signify an interest in the design of mobile applications that can track social interactions, and deliver tailored services to the end user. 

\paragraph{Effects of physical spaces on social interactions}
Those who plan and design buildings have long been interested in the `social logic of space'~\cite{Hillier84:Social}, that is, how the nature and layout of spaces can affect patterns of usage and group behavior.  One set of methods to aid such study is those of \textit{space syntax}, used by Penn \emph{et al.}~\cite{Penn99:Space} in a 1999 study finding that the spatial configuration of a work environment can directly affect the frequency of face-to-face communication between office workers. Furthermore, the frequency of such contact affected how useful employees found work-related interactions to be. The same importance of the role of space in the frequency of communication between workers was also found in a later study by Sailer and Penn~\cite{sailer2009spatiality}.

In the same vein, Toker and Gray~\cite{Toker08:Innovation} studied the effect of workspace layout specifically in the context of research offices and laboratories, such as that that we study here, and found that spatial configuration affected the frequency and location of unplanned face-to-face interaction between workers.

\paragraph{Cultural dimensions}
Beginning in 1971, Geert Hofstede conducted an extensive study of the way that cultural differences shape the way that people think, feel, and act, and can therefore influence workplace environments and office social dynamics. He administered over 100,000 opinion surveys to IBM employees in 70 countries, and through his analysis derived five \textit{cultural dimensions} that could explain most of the variation between cultures in his observations~\cite{Hofstede10:Cultures}.

These factors were recently shown, in a study by Reinecke \emph{et al.}~\cite{Reinecke13:Doodle}, to correlate with differences in behavior of office workers in different countries scheduling meetings, which demonstrates that cultural differences continue to have an effect on workplace behavior especially in today's highly international business world.

The idea of Hofstede's cultural dimensions was also recently applied in the context of office space layout by researchers from the office furniture company Steelcase~\cite{Steelcase13:Culture}. Over a period of 5 years, they studied offices in 11 different countries including China, India, Italy, Germany, and Britain, and showed how differences in national culture could mean that difference office layouts would be more effective in different countries. For example, in countries where competition tends to be valued over collaboration, private offices are important and collaboration spaces may be very basic, while in countries placing higher value on cooperation, workers may be helped by more open, fluid spaces.

In this study, we show that besides cultural differences between countries where offices are situated, in a highly international work environment there may also be such effects at work in a single office.

\section{Measuring serendipitous interactions}
We study a culturally diverse environment, populated by workers from many different backgrounds. We have two aims: firstly, to examine how cultural differences, as measured by Hofstede's cultural dimensions, may affect the serendipitous interactions of individuals. Secondly, we wish to identify the spaces within the workplace that are most likely to host such interactions. Our approach is based on the collection of real-world traces of social interactions in a culturally diverse research institution, then exploring how cultural dimensions are related to serendipitous interactions, through quantitative analysis.

\subsection{Cultural dimensions and interactions of individuals}
As a result of his study of cultural variations across countries in the 1970s, Hofstede defined five \textit{cultural dimensions} that could explain most of the variance in his data~\cite{Hofstede10:Cultures}. In this work, we study three of these\footnote{The remaining two, Uncertainty Avoidance and Long-Term Orientation, are less directly associated with short-term interactions between people as considered by our study. For example, uncertainty avoidance is often reflected in whether company managers are focused on day-to-day operations, which demand less tolerance of uncertainty, or on strategic problems, which are by nature more uncertain. Long-term orientation is relevant in situations such as businesses deciding whether to focus on short-term profits or future growth~\cite{Hofstede10:Cultures}.}:

\begin{itemize}
	\item \textbf{Power Distance} describes how comfortable people tend to be with unequally distributed power and a clear social hierarchy. In a typical office environment the distribution of power is reflected by the different roles that are defined in the workplace, and the explicit or implicit hierarchy across them.  
	Given Hofstede's definition of Power Distance, we consider that people less comfortable with differences in power may be less likely to interact with those with colleagues with a different role from their own, and propose the hypothesis:\\ \emph{HP1: Individuals from cultures with lower Power Distance tend to have a lower proportion of their interactions with those outside their own roles.}
	\item \textbf{Individualism} concerns the extent to which self-sufficiency is valued and people largely look after only themselves (high Individualism), or whether group well-being is important and people have more responsibility for others. Collaborations in many office environments are commonly carried out by clearly defined groups consisting of people in a variety of roles, where all members work together to achieve a common goal. 	We consider that people from more collectivist cultures (lower Individualism) may perceive interactions with others in different roles important to maintaining group well-being, and form the hypothesis:\\\emph{HP2: Individuals from cultures with lower Individualism tend to have more of their interactions with those outside their own roles.}
	\item \textbf{Masculinity} quantifies how task-oriented (as opposed to person-oriented) a culture tends to be. Cultures with high Masculinity place higher value on achievement and competition, while cultures with lower Masculinity regard relationships and co-operation as more important. We consider that people from more person-oriented and less task-oriented cultures may regard a person's work role as less relevant to their willingness to interact, and test the hypothesis:\\\emph{HP3: Individuals from cultures with lower Masculinity tend to have more of their interactions with others outside their own roles.}
\end{itemize}

In this work we attempt to answer these research questions experimentally, through the collection of real-world traces of social interactions in a culturally diverse office environment.

\section{Data collection}
\begin{figure*}[t]
\includegraphics[width=\columnwidth]{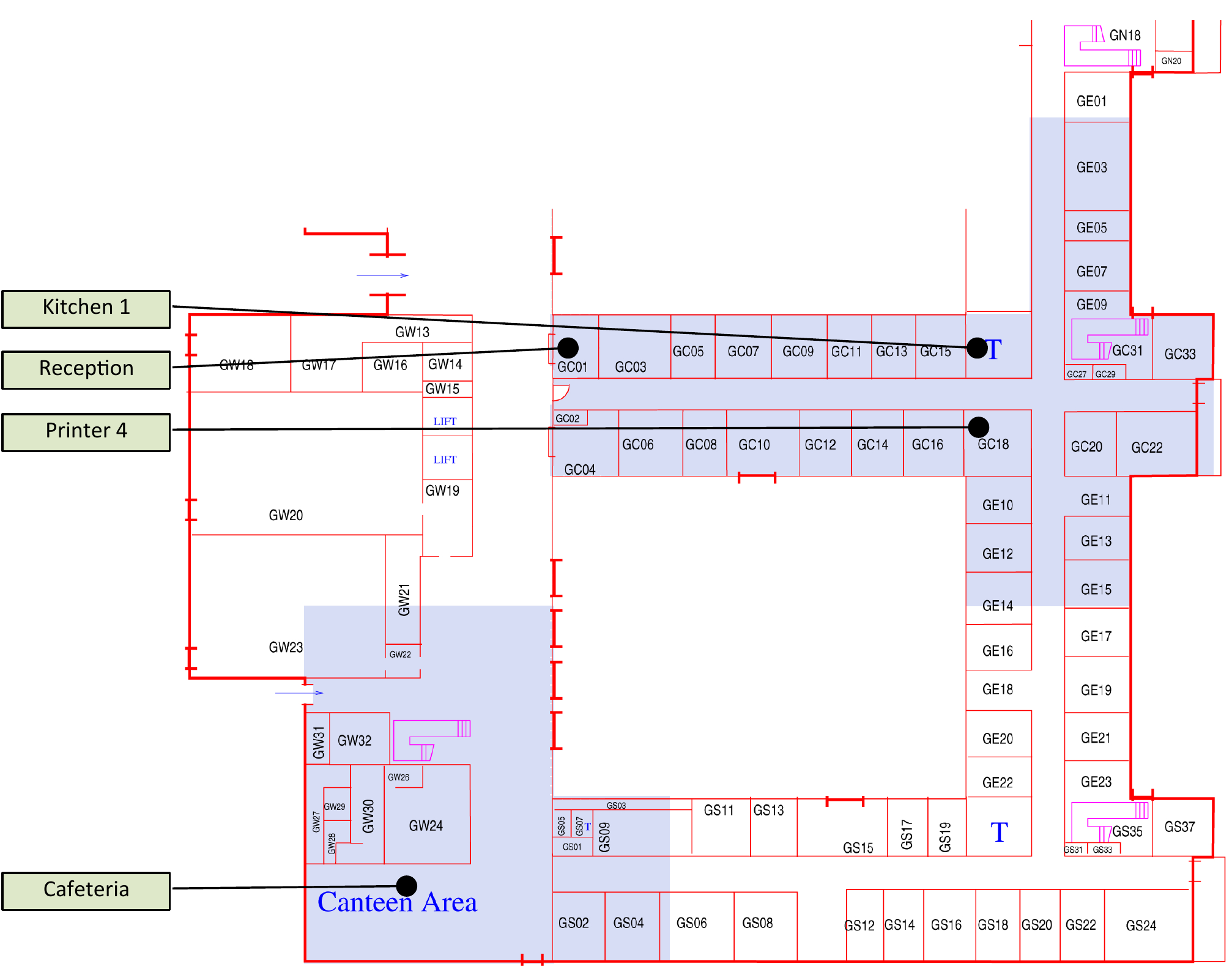}\hfill
\includegraphics[width=\columnwidth]{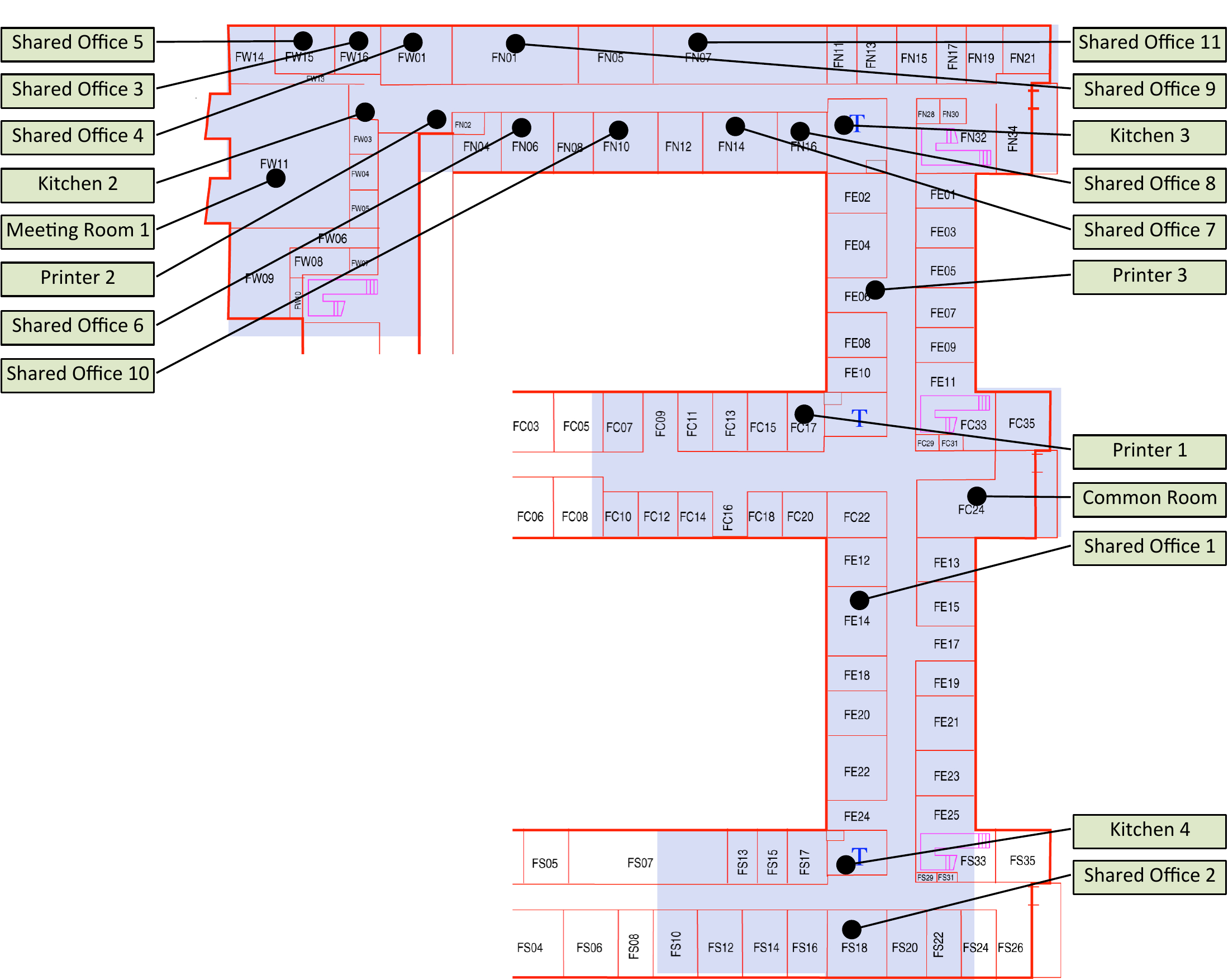}
\caption{Data was collected on the  shadowed areas of the ground floor (left) and the first floor (right) in a UK research laboratory. }
\label{fig:map}
\end{figure*}

\begin{figure}
\centering
\includegraphics[width=.9\columnwidth]{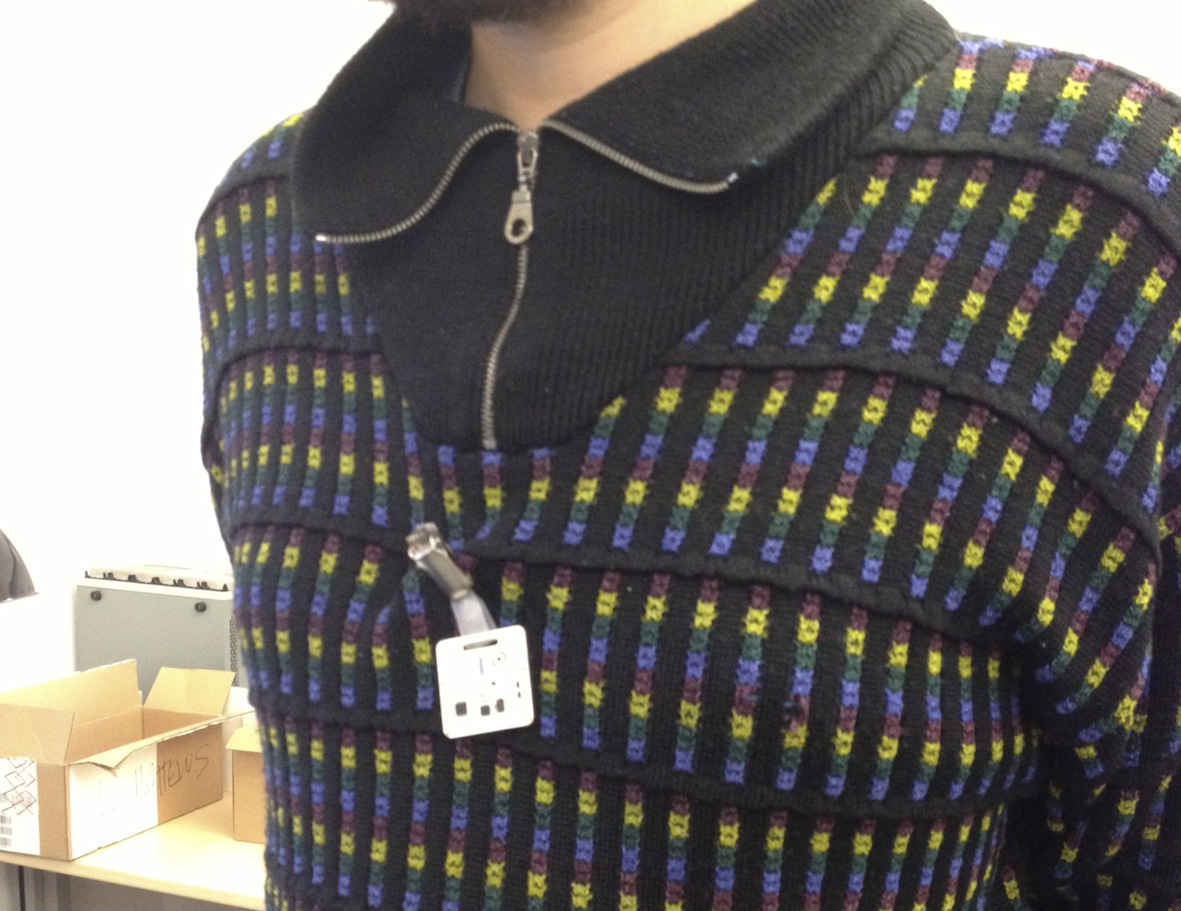}
\caption{The SocioPatterns tag worn on the chest is able to track face-to-face interactions.}
\label{fig:openbeacon}
\end{figure}

We conducted a study over 2 weeks in 2012 within a research laboratory in the UK. We collected traces of face-to-face interactions, using the SocioPatterns proximity-sensing platform, which was made available to us
in order to carry out the measurements. The measurements were captured by active RFID badges~\cite{Cattuto10:Dynamics}, worn on the body as shown in Figure \ref{fig:openbeacon}. The badges are lightweight radio transceivers,  programmed to transmit a low signal strength beacon periodically, and to listen continuously for beacons from other badges nearby. 
The badges are configured to transmit low signal strength beacons that were experimentally evaluated to have a range of 1.5m - 2m with clear line-of-sight. When worn by the participants, the beacons are shielded by the body, meaning that successful communication can occur only when another badge is facing that of the participant. This way the tags can assess continued face-to-face proximity between users. We assume continued face-to-face proximity to be a good proxy for a social interaction between users. Defining the threshold for such matching to be 2m (the maximum range of the radio transmission) makes the likelihood of false positives in the dataset negligible. Reducing the number of false negatives (face-to-face proximity not detected by the tags) can be controlled by using time windows within which detected beacons can be considered as indicators of proximity for that duration. In the study by Panisson \emph{et al.}~\cite{Panisson11:Dynamics} the authors established that the use of a 20-second window offers a 99\% probability for a face-to-face proximity detection.

We also augmented the office environment with a number of active RFID tags placed on the walls of rooms. The 26 rooms with static tags on the walls were: 4 kitchens, the common room, the cafeteria, reception, 13 offices, 4 printers, and 2 meeting rooms (Figure~\ref{fig:map}). The room tags were programmed to beacon a unique location ID at a higher signal strength than the participants' badges, achieving a range of 5m - 6m.

Traces from both badge-to-badge interactions, and from contacts between badges and location tags, were collected by RFID readers installed throughout the building. This instrumentation allowed us to capture two sets of data: the first consisting of timestamped face-to-face contacts between participants, and the second involving proximity of participants to particular locations within the building.

Study participants were recruited based on the physical locations of their offices, with the aim of including the majority of people working within a specific geographic space (the north wing of the main laboratory building). 
After briefing all employees working within the target area, we recruited 61 participants, which represents more than 80\% of the total number of employees in the participating areas.

\begin{figure}[t]
\centering
\includegraphics[width=\columnwidth]{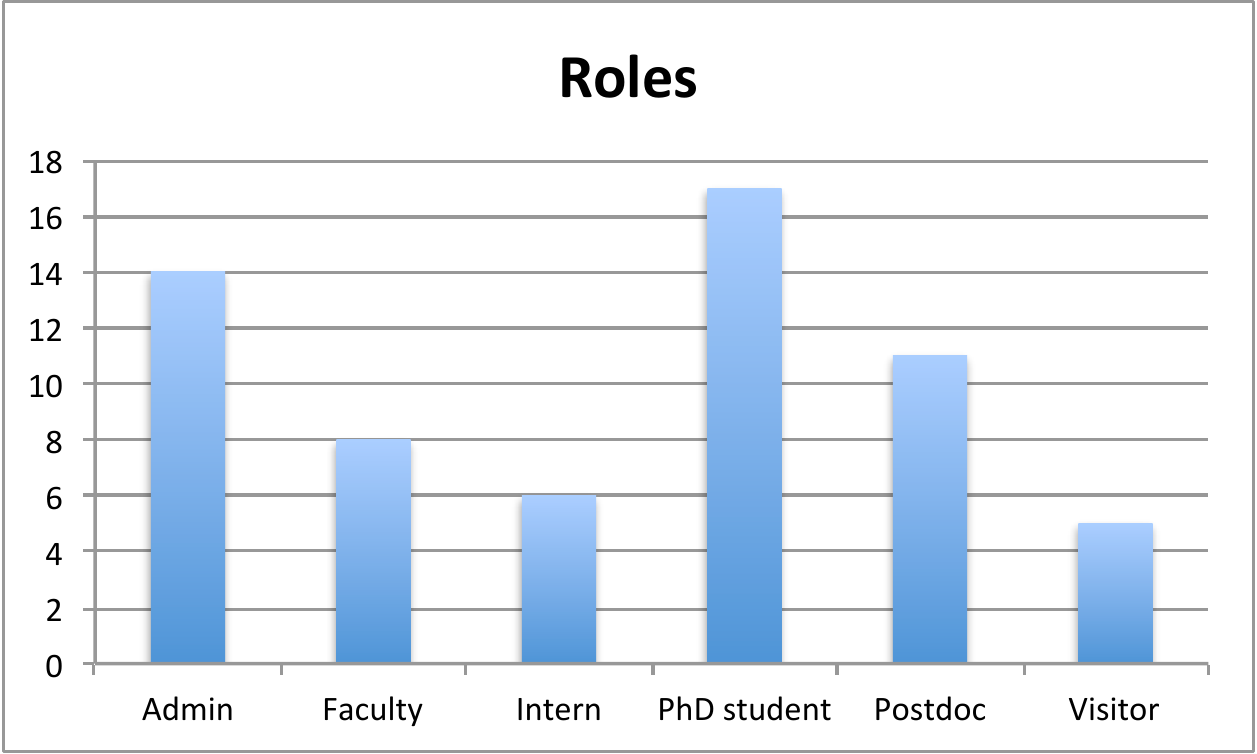}
\caption{Distribution of roles across the study participants.}
\label{fig:userroles}
\end{figure}

\begin{figure}[t]
\centering
\includegraphics[clip=true,trim=0pt 14cm 0pt 0pt,width=\columnwidth]{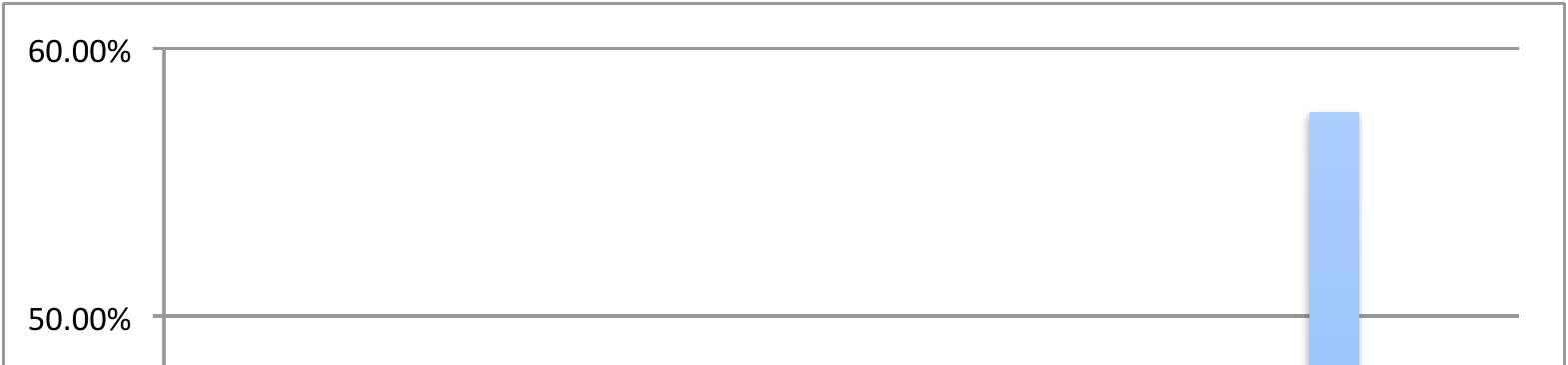}
\includegraphics[width=\columnwidth]{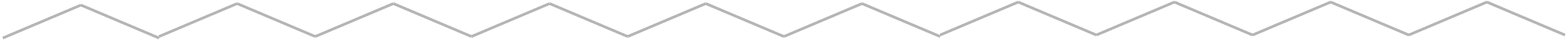}
\includegraphics[clip=true,trim=0pt 0pt 0pt 12cm,width=\columnwidth]{figures/nationalities.pdf}
\caption{Distribution of nationalities across the study participants.}
\label{fig:nationalities}
\end{figure}

At the start of the 2-week data collection period, the participants completed a survey indicating their job role (e.g.,~PhD student, postdoc, lecturer), their working group (team), their age, and their gender. They also identified among the other participants people with whom they collaborate on work projects, and people they consider to be their friends.
The demographics of the participants reflect the overall demographics of the particular research institution. The participant group was 78.6\% male and 21.3\% female, with age distributed as 34.4\% 20--30, 22.9\% 30--40, 14.7\% 40--50, 21.0\% over 50, and 4.9\% not specified. The participants were spread across different roles as shown in Figure~\ref{fig:userroles}. 
With respect to different nationalities, as expected, the group was dominated by UK nationals (Figure~\ref{fig:nationalities}). However, UK nationals made up only 57\% of the total population, making this a diverse working environment where people from different cultural backgrounds interact on a daily basis.

Over the 2 weeks of the deployment, the badges and tags generated over 270,000 face-to-face contact reports, and over 730,000 location proximity traces.

\begin{table}
\small\centering
    \begin{tabular}{|l|c|c|c|}
    \hline
    \textbf{Country} & \textbf{Power Dist.} & \textbf{Individualism }& \textbf{Masculinity}\\\hline
Chile & 63 & 23 & 28\\\hline
China & 80 & 20 & 66\\\hline
Germany & 35 & 67 & 66\\\hline
Greece & 60 & 35 & 57\\\hline
India & 77 & 48 & 56\\\hline
Italy & 50 & 76 & 70\\\hline
Norway & 31 & 69 & 8\\\hline
Pakistan & 55 & 14 & 50\\\hline
Russia & 93 & 39 & 36\\\hline
Spain & 57 & 51 & 42\\\hline
UK & 35 & 89 & 66\\\hline
USA & 40 & 91 & 62\\\hline
    \end{tabular}
        \caption{Hofstede's cultural dimensions of the participants' nationalities. Each dimension can have a value from 1 to 120.}
    \label{tab:dimensions}
\end{table}

\subsection{Individual interactions}
Hofstede assigned each of the countries he studied a score from 1 to 120 for each dimension (Table \ref{tab:dimensions}). To test our hypotheses, we therefore assign to each study participant the score for their self-reported country of origin, for each dimension\footnote{Note that we did not administer Hofstede's original questionnaire, as this would have risked biased results due to participants altering the way they interacted, being overly aware of their behavior in relation to the cultural dimensions.}. We then compute for each individual their \textit{Interaction Diversity}, $ID$, defined for each participant $p$ to be:
\begin{equation}
ID(p) = \frac{\text{\#people $p$ interacted with in different role to $p$}}{\text{\#people $p$ interacted with}}
\end{equation}

We then test our hypotheses by examining the relationship between individuals' Interaction Diversity and their scores for each of the three cultural dimensions.

\subsection{Serendipitous spaces}
We also study the nature of different kinds of space within the building, aiming to characterize their potential to host encounters between people from different teams or groups. Static tags were placed on the walls of shared offices, individual offices, kitchens, printer rooms, and a common room and cafeteria, allowing the occupants of the room to be detected by means of the signals from their badges. 

Specifically, for each room, we compute the diversity of the visitors to that room by using the Shannon index $H'$, a measure often used in ecology to measure the diversity of populations, and originally proposed by Claude Shannon to quantify the entropy in strings of text~\cite{Shannon48:Mathematical}.

The index is defined as:
\begin{equation}
H' = -\sum_{i=1}^{R}p_i \ln p_i
\end{equation}

where $p_i$ is the proportion of visitors to the room belonging to group $i$ out of $R$. The higher the Shannon index, the more diverse the population of visitors to a room and, as such, the more potential the room has for hosting serendipitous interactions between members of different groups. If all the visitors to the room are from the same group, $H'$ has a value of 0.

We measure $H'$ based on several different definitions of `group', these being: job role, research group (team), gender, age, and nationality. We aim to answer two questions:
\begin{enumerate}
	\item \emph{Which places are visited by people from many different groups, and therefore likely places for serendipitous meetings?} This question is motivated by the statement in~\cite{Isaacs96:Piazza} that cafeterias and photocopiers are particularly likely places for such encounters.
	\item \emph{Are the same places the most diverse across all definitions of `group', or are some more diverse in some terms than in others?} Serendipitous meetings between those from different groups may be beneficial for the exchange of ideas and information whether those encountering one another are from different social circles, different teams, or have different job roles and therefore different perspectives and expertise. The answer to this question could provide insight into which kinds of places are the best for fostering serendipitous interactions between people from different groups depending on the type of group being considered.
\end{enumerate}

\subsection{Pre-processing}
We pre-processed the data collected by the RFID badges in order to extract locations of participants in rooms, and face-to-face contacts between participants.

To record a person as visiting a room, for the purpose of computing diversity in that room as described above, we required that there be a contact between that person's badge and the static badge in the room. We further required that more than 90\% of the contacts between the person's badge and static badges in rooms be from the room in question, over a 30-second time window. The reason for this window was to remove noise from events where, for example, people were walking around and were `seen' by multiple room badges over a short period of time.

To record a face-to-face contact between two people, we aggregated all contacts between the same two badges with an interval less than 30 seconds in between into the same interaction, resulting in a total of 4646 interactions being recorded. In the analysis that produced the following results, we considered these aggregated interactions.

\section{Results}
\subsection{Cultural dimensions and interactions of individuals}
We now present the results of testing the hypotheses stated earlier. For each cultural dimension we compute Pearson's correlation coefficient $r$ and the corresponding $p$-value. The descriptions of the cultural dimensions are given again in Table \ref{tab:cultural} for easy reference.

\begin{table}[t]
    \begin{tabular}{|p{2.8cm}|p{5cm}|}
    \hline
    \vskip 1em \textbf{Power Distance} & How comfortable people tend to be with unequally \mbox{distributed} power and a clear social \mbox{hierarchy}                                                                    \\\hline
    \vskip 2em \textbf{Individualism}
      & How much self-sufficiency is \mbox{valued} and people largely look after only themselves, or group well-being is important and \mbox{people} have more responsibility for others \\\hline
     \vskip .5em  \textbf{Masculinity}
       & How much achievement and \mbox{competition} are valued, as \mbox{opposed} to relationships and co-operation                                                                  \\\hline
    \end{tabular}
    \caption{Descriptions of cultural dimensions}
    \label{tab:cultural}
\end{table}

\emph{HP1: Individuals from cultures with lower Power Distance tend to have a lower proportion of their interactions with those with different roles.}
Table \ref{tab:results} shows a positive correlation between Power Distance and Interaction Diversity, as defined earlier. When all contacts are considered, $r=0.31$, confirming that individuals from cultures with lower power distance do indeed have a lower proportion of their interactions with those having roles different from their own. Considering only contacts with those the study participants identified as collaborators, the correlation is stronger ($r=0.36$), and weaker considering only contacts with those identified as friends ($r=0.25$). If only interactions with those in a different team are considered, the result is much the same as that where all contacts are considered ($r=0.32$, as opposed to $r=0.31$), so the relationship appears to be unaffected by whether workers are on the same team or not. These results effectively mean that individuals from cultures where there is lower acceptance of inequality of power are less likely to interact with those above or below them in the workplace hierarchy, and that this effect is strongest for working relationships.
\begin{table*}[tbh]
\begin{centering} 
\begin{tabular}{l|r|r|r|r|}
\cline{2-4}
& \multicolumn{1}{ c| }{Power distance}&\multicolumn{1}{ c| } {Individualism}&\multicolumn{1}{ c| }{Masculinity} \\\hline 
\multicolumn{1}{ |l| }{\multirow{1}{*}{All contacts} } &  \textbf{0.31} (0.02) &  \textbf{-0.27} (0.04) &  \textbf{-0.27}  (0.04)\\ \hline
\multicolumn{1}{ |l| }{\multirow{1}{*}{Collaborators} } &  \textbf{0.36} (0.01) &  \textbf{-0.32} (0.03) &  \textbf{-0.36} (0.01)\\ \hline
\multicolumn{1}{ |l| }{\multirow{1}{*}{Friends} } & 0.25 (0.13) & -0.23 (0.17) & \textbf{-0.35} (0.03)\\ \hline
\multicolumn{1}{ |l| }{\multirow{1}{*}{Inter-team only} } &  \textbf{0.32} (0.04) &  \textbf{-0.31} (0.04) & -0.27 (0.08)\\ \hline
\end{tabular}
\caption{Pearson's correlation coefficient $r$ (and $p$-value) for the three cultural dimensions under study  vs.~Interaction Diversity.}
\label{tab:results}
 \end{centering}
\end{table*}

\emph{HP2: Individuals from cultures with lower Individualism tend to have more of their interactions with those with different roles.}
Table \ref{tab:results} shows a negative correlation between Individualism and Interaction Diversity, which confirms that those from cultures with lower Individualism, where the group is considered to be important (rather than the self alone), tend to interact more with those in different roles from themselves. Again, the correlation considering interactions with collaborators ($r=-0.32$) is slightly stronger than that obtained from considering all interactions ($r=-0.27$), which suggests that this effect may be particularly important in the context of working relationships and less relevant for purely social relationships. If only interactions with those in a different team are considered, the result is much the same ($r=-0.31$) as that where all contacts are considered, so the relationship appears to be unaffected by whether workers are on the same team or not. Individuals from cultures with low Individualism may perceive that interacting with their colleagues having different roles is important for the wellbeing of the group as a whole, to a greater degree than those from backgrounds where self-sufficiency is valued more highly.

\emph{HP3: Individuals from cultures with lower Masculinity tend to have more of their interactions with others outside their own roles.}
Table \ref{tab:results} shows a negative correlation between Masculinity and Interaction Diversity, in agreement with the hypothesis: those from more task-oriented cultures, where achievement and competition are more valued, interact less with those in different roles than those from more person-oriented cultures, where co-operation and relationships are more valued. As for the other two cultural dimensions, the correlation is stronger for working relationships ($r=-0.36$) than when considering any relationship ($r=-0.27$). For Masculinity in particular, there is also a stronger correlation ($r=-0.35$) when considering only interactions with those the participant in question identified as friends. If only interactions with those in a different team are considered, the result is much the same as that where all contacts are considered ($r=-0.27$), although the $p$-value in that case is 0.08, rather than 0.04. People from cultures with low Masculinity are more likely to engage in interaction with those in different roles from themselves, but the effect is equally strong for working relationships and social relationships -- those with a less task-oriented and more person-oriented background may be more open to social interaction with those in different roles from themselves. 

\subsection{Serendipity of rooms}
The results of computing the diversity of visitors to rooms according to five different ways of defining groups (role, research group, gender, age, and nationality) are given in Table \ref{tab:rooms}. To facilitate understanding of these results, we should remember that the two floors under study consisted of a variety of places where diverse people could potentially meet, as shown by Figure~\ref{fig:map}. The ground floor included reception, administrative offices, a printer, a kitchen, and one big cafeteria, and the first floor included  three printers and three kitchens shared by different research groups, offices shared  by PhD students and postdocs, and one common room. The table shows the top 10 locations with the most diverse range of visitors, according to the Shannon index $H'$ computed over the various kinds of group.
{

\begin{table*}[tbh]
   \small
    \begin{centering}
    \begin{tabular}{|c|lr|lr|lr|lr|lr|}
    \hline
    & \multicolumn{2}{|c}{Role}           & \multicolumn{2}{|c}{Research group} & \multicolumn{2}{|c}{Gender}          & \multicolumn{2}{|c}{Age}         & \multicolumn{2}{|c|}{Nationality}     \\\hline
    1    & Printer 2& 1.79       & Shared office 1& 1.72 & Kitchen 1& 0.63       & Reception & 1.61  & Shared office 3 & 2.04\\\hline
    2    & Common room  & 1.79   & \textbf{Printer 1}   & \textbf{1.71}    & Printer 4   & 0.57    & Cafeteria & 1.58  & Kitchen 2   & 2.02    \\\hline
    3    & Shared office 3& 1.77 & \textbf{Shared office 2} & 1.67 & Reception  & 0.56     & Common room& 1.57 & Meeting room 1& 2.02   \\\hline
    4    & Shared office 1& 1.77 & Common room  & 1.64   & \textbf{Shared office 4}& \textbf{0.54} & Printer 4 & 1.57  & \textbf{Shared office 5} & \textbf{1.94}\\\hline
    5    & Printer 3  & 1.77   & Kitchen 1  & 1.63  & Cafeteria  & 0.52     & Printer 3 & 1.56  & Shared office 6&  1.93\\ \hline
    6    & Cafeteria  & 1.76   & Reception  & 1.63  & Common room  & 0.51     & Kitchen 1 & 1.54  & \textbf{Shared office 11}& \textbf{1.92} \\ \hline
 7    & Kitchen 2  & 1.76   & Printer 4  & 1.63  & Shared office 6  & 0.50     & Kitchen 3 & 1.53 & \textbf{Shared office 8}& \textbf{1.92} \\ \hline
 8    &  Kitchen 3 & 1.75   & Cafeteria  & 1.63  & Printer 2  & 0.45     & Printer 2 & 1.47  & \textbf{Shared office 9}& \textbf{1.92} \\ \hline
 9    & Meeting room  1& 1.72   & \textbf{Kitchen 4}  & \textbf{1.63}  & Kitchen 2  & 0.41     & Shared office 7 & 1.41  & Shared office 7& 1.90 \\ \hline
 10    & Reception  & 1.71   & Printer 3  & 1.63  & Shared office 3  & 0.41     & Meeting room 1& 1.36  & \textbf{Shared office 10}& \textbf{1.86} \\ \hline
    \end{tabular}
    \caption{Top 10 most diverse rooms and the corresponding values of $H'$, for each way of defining groups. Bolded entries indicate rooms that do not appear in the top 10 for any other definition of groups.}
    \label{tab:rooms}
    \end{centering}
\end{table*}
}

With regard to our first question posed earlier:\\
\emph{1. Which places are visited by people from many different groups, and therefore likely places for serendipitous meetings?}
We can see that all 4 printers feature in at least one of the lists, and all the lists except that for Nationality feature at least two printers. This would support the idea that printers are likely places for serendipitous, inter-group encounters. Similarly, at least one of the social spaces including the kitchens, cafeteria, and common room appears among the top 3 places on each of the lists, and all of the lists except that for Nationality include two kitchens. This would also suggest that kitchens are indeed likely places for serendipitous meetings between people from different groups.

We can also consider the second question:\\
\emph{2. Are the same places the most diverse across all definitions of `group', or are some more diverse in some terms than in others?}
The bolded entries in the table show that there are indeed some rooms that are highly-ranked only in one list, and all but 2 of these places are shared offices. It seems likely that this is because while kitchens and printers are, by nature of the usage of the space, ideal for mixing between people from different groups, shared offices are often occupied by people from the same research group and having the same role. These results meet what one would expect, and that speaks to their external validity and to the quality of our experimental design. 

The Nationality list is noticeably different in that most of the entries in the list are offices. This can be regarded as a reflection of the highly culturally diverse environment of the research lab as a workplace. Offices are populated by many different people from many different cultural backgrounds, and this highlights the relevance of the relationships between Hofstede's cultural dimensions and social interaction patterns in the workplace that we presented above.

\section{Discussion}
From a theoretical standpoint, this study provides a modern perspective on an established set of metrics that are commonly used to explain inter-culture interaction. Our work shows how new technology, such as the sensing badges that we have deployed here, can be used alongside well-established theory, and our results provide evidence in support of the validity of these metrics and theories.

In practical terms, we have demonstrated that the way that space in the workplace is used can indeed result in more or less diversity, according to a variety of definitions, in certain locations. Our work paves the way for new application opportunities, such as better tools to understand and visualize the cultural interactions within an organization.

\textbf{Theoretical implications.} Past experimental work has suggested that ideas for productive collaboration will most likely come from `idea brokers', those who maintain broad networks across many company divisions~\cite{fishman13better,krack93informal}. Furthermore, that the best-performing groups seek fresh perspectives by frequently interacting with other groups~\cite{Olguin09:Sensible}. Informal interactions can foster serendipitous mixing of ideas, which fuels innovation, but these interactions evolve so rapidly that they are not easy to track. Using the active RFID tags, we have been able to track them and validated the corresponding serendipity metric: it is highest in spaces in which it is expected to be so, which suggests that the badges are good for performing this type of study. In contrast to previous technological solutions like collar devices, the badges are much less obtrusive (they are small, barely noticeable,  and lightweight) and can capture data at scale (they allow for tracking a large number of participants). They are able to observe, quantify, and measure interaction dynamics that are often associated with productive teams. Previous work has found that the most valuable form of communication is face-to-face (``35\% of the variation in a team's performance can be accounted for simply by the number of face-to-face exchanges among team members''~\cite{Pentland12:New}), while the least valuable forms are e-mail and mobile text messages.

The significance of these results extends beyond merely tracking interactions: they provide evidence that not only the culture of an office, but also the cultures of individuals within an office, could shape interactions. To appreciate the importance of this insight, consider that past research has focused on offices as self-contained entities in the countries where they are situated~\cite{Steelcase13:Culture,Hofstede10:Cultures}, but not on how cultural differences between individuals in highly international work environments can shape office social dynamics. In addition, our research makes key informal dynamics measurable to an unprecedented degree, producing findings based upon highly reliable observations.

\textbf{Practical implications.}
Our results show that the way that space in the workplace is used can indeed result in more or less diversity in certain locations, depending on the kind of diversity that it is desirable to achieve. From the location results, we can see that some shared offices are ranked highly for diversity along certain dimensions (e.g.,~gender or nationality) and appear in no other list. This is because in the lab in question, shared offices are often occupied by people from the same research group and having the same role (e.g.,~offices occupied entirely by PhD students with the same supervisor). Therefore, these offices will not promote interactions between people from different roles or research groups, but will have a diverse population along dimensions not closely related to role or research group. It is true that the assignment of people to offices does not completely preclude diverse interactions, since others may visit the office. However, these interactions may more commonly take place in other spaces, to avoid disturbing the other occupants of the office.

On the other hand, other areas such as kitchens and printers are located more centrally, such as between corridors used by different research groups, and may therefore be visited by people from both groups. Similarly, common areas such as these are not used exclusively by people of one or two roles, e.g., a single printer or coffee machine will be used by faculty and students alike, so there is more diversity in these areas along dimensions such as role.

To the best of our knowledge, this is the first investigation into workplace spaces, culture, and types of interaction conducted using purely automatic mechanisms.  This has implications for future such studies, in showing how workplace interactions could be studied in conjunction with new physical workplace configurations.

Our findings might translate into a number of practical applications. To give one example, we are currently planning to  create maps of the lab's floor plans reflecting the extent to which each area hosts serendipitous conversations, and graphs showing how a research group is doing in terms of serendipitous interactions. If these maps were to be made accessible, they would provide instant visual feedback to anyone, and their impact on group interactions could be measured. In the future, one could imagine a lab's entire staff wearing our name badges, creating what Pentland calls a `God's-eye view' of the organization~\cite{Pentland12:New}.

\textbf{Limitations.} This study has three limitations that call for further investigation in the future. The first is that our results do not speak to causality. While we have found that Hofstede's cultural dimensions of Power Distance, Individualism, and Masculinity are correlated with interactions with others across different roles, the root cause of this is not clear from this study, and there would therefore be benefit to be gained from more studies of a similar nature.

Secondly, we did not compare serendipitous interactions to any performance metric. In the research lab, it is difficult to find a well-defined performance metric apart from publications and research impact, which do not change frequently enough to warrant real-time tracking of informal interactions. However, this does not mean that such interactions do not have an important effect, just that it is difficult to measure formally using existing metrics. As a next step, we are planning to collect subjective daily measures (e.g., frustration, satisfaction) from our participants using experience sampling, and to investigate any relationship between these measures and sensed serendipitous interactions.

A third limitation is that we have not studied exactly how the sensing of employees' inter-group interactions, and meetings in serendipitous spaces, could be used in practice by an organization, for example, to foster more such interactions. One could envisage that making workers aware of the sensed information could prompt them to consider their own interactions and might have a positive effect on productivity by increasing the exchange of information and ideas. 

\section{Conclusions}
In this work, we have used state-of-the-art active RFID tags to track serendipitous interactions in the workplace between individuals from different groups, in a less obtrusive and therefore potentially more accurate way than was possible using previous sensing technology. 

Our results suggest that cultural differences between individuals in a highly international office environment can affect the likelihood that people engage in serendipitous interactions with others from different groups.  We have also been able to characterize certain spaces in the workplace, such as kitchens and printers, as being particularly likely to host serendipitous interactions, and shown that different kinds of spaces may be more likely to host inter-group interactions, depending on the kind of group considered. Knowledge of these effects could be important for organizations, given that such interactions have been shown by previous work to be beneficial for group productivity, by enabling the exchange of information and ideas.

\section{Acknowledgements}
We thank the ISI Foundation and Bitmanufaktur for their support. We acknowledge the SocioPatterns collaboration (\texttt{http://www.sociopatterns.org}). We also thank Jon Crowcroft, and Kerstin Sailer for valuable discussions and feedback. Chlo\"e Brown is a recipient of the Google Europe Fellowship in Mobile Computing, and this research is supported in part by this Google Fellowship.

\bibliographystyle{ieeetr}
\bibliography{serendipity-arxiv}
\end{document}